# One-directional polarization transport in electron/nuclear spin chains with loss and gain

Santiago Bussandri[1,2], Pablo R. Zangara[1,2], Rodolfo H. Acosta[1,2], and Carlos A. Meriles[3,4,*]

[1]*Universidad Nacional de Córdoba. Facultad de Matemática, Astronomía, Física y Computación, Córdoba, Argentina.*
[2]*CONICET. Instituto de Física Enrique Gaviola (IFEG), Córdoba, Argentina.* [3]*Department of Physics, CUNY-City College of New York, New York, NY 10031, USA.* [4]*CUNY-Graduate Center, New York, NY 10016, USA.*

[*]*Corresponding author. E-mail: cmeriles@ccny.cuny.edu.*

Understanding the joint dynamics of electron and nuclear spins is central to core concepts in solid-state magnetic resonance — such as spin-lattice relaxation and dynamic nuclear polarization — but a generalization that capitalizes on competing polarization loss and gain channels is still lacking. Here, we theoretically study the non-Hermitian dynamics of hybrid electron/nuclear spin systems in the simultaneous presence of electron spin pumping and spin-lattice relaxation. Focusing on periodic, one-dimensional chains, we find that by adjusting the electron spin pumping to a critical level, it is possible to steer the flow of nuclear polarization to create site-dependent distributions where either end of the array polarizes in opposite ways, irrespective of the initial state. By contrast, we show that ring-like patterns — where the limit nuclear polarization is uniform — exhibit a non-decaying, externally-driven nuclear spin current. Interestingly, cyclic magnetic field modulation can render these processes largely robust to defects in the chain, a response featuring some interesting similarities — and differences — with recent findings in other non-Hermitian physical platforms.

## I. INTRODUCTION

Since any realistic quantum system interacts with its environment, a viable control protocol must ensure that 'dissipation' channels created by unwanted couplings be eventually compensated by 'gain' processes able to re-introduce order. The development of the (necessarily) non-Hermitian formalism describing this interplay has gradually led to the discovery of a range of phenomena not anticipated for isolated, Hermitian systems. The clearest illustrations can be found in the investigation of photonic platforms, where various practical applications of non-Hermiticity are currently being explored including enhanced sensing[1], mode-selective laser cavities[2,3], unidirectional invisibility[4], or loss-induced transparency[5]. Remarkably, recent work in a photonic mesh with tailored anisotropy of the nearest-neighbor coupling has shown that it is possible to make all optical modes — otherwise extended through the lattice — localize at the interface between regions of the mesh with different topologies[6].

Since non-Hermiticity inherently leads to the breaking of time-reversal symmetry, a large effort is being devoted to exploring the connection with non-reciprocity, i.e., the invariance of a system upon exchange of the emitter and receiver. In particular, it has been shown that it is possible to induce one directional photon propagation by resorting, e.g., to non-linearity[7-9] or to spatiotemporal modulation[10,11]. These strategies recreate the response observed in Faraday rotation and other related magneto-optical phenomena without the need for an externally applied magnetic field[12]. Beyond photonics, similar techniques have been adapted to areas such as acoustics[13], where non-reciprocal transport is otherwise difficult to attain.

While a large fraction of the above work draws on the ability to map the system's fundamentally classical dynamics into a Schrödinger-like equation, less attention has been devoted to physical settings in the form of interacting nuclear and electron spins, where a quantum mechanical description is typically mandatory. Further, with few exceptions[14], we are not aware of any studies considering the combined effect of coexisting spin polarization gain and loss, even if the impact of spin-lattice relaxation on the system's time evolution is well understood[15]. At first sight, this void may seem surprising because dynamic nuclear polarization (DNP) — the process whereby external optical or microwave (MW) excitation leads to nuclear spin order[16] — has been known for decades. Partly to blame is the very different time and frequency scales governing electron and nuclear spin dynamics, and thus the entrenched notion of electron spins as a generic source of nuclear spin relaxation in solids, typically cast in a semi-classical — and thus uninformative — fashion.

Here, we theoretically consider a hybrid set of nuclear and electron spins simultaneously undergoing optical spin pumping and spin-lattice relaxation. We zero in on chain-like arrays where the nuclear spin coupling is mediated by pairs of interacting electron spins, one of which spin-polarizes under laser excitation. Setting the externally applied magnetic field so as to allow inter-electronic spin flips, we show that when the spin pumping rate reaches a critical threshold, nuclear spin polarization flows selectively in one direction of the chain but not in the other, a response reminiscent of a spin valve. The dynamics that ensues leads to asymmetric, site-selective nuclear spin polarization along the chain, which we formally capture with the help of an effective nuclear-spin-only non-Hermitian model. Transitioning from linear spin chains to ring-like arrays, we identify a drastically different limit regime, this time featuring uniform polarization distributions as well as persistent nuclear spin currents, even in the absence of net nuclear polarization. Finally, we show that, to a large extent, these processes can be made robust to defects in the chain through the periodic modulation of the applied magnetic field.

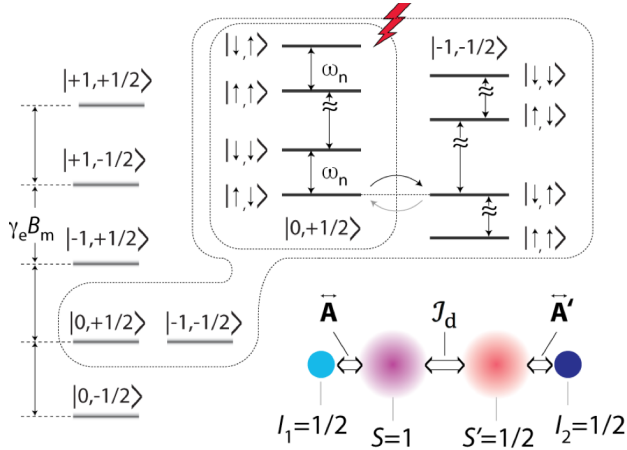

**Fig. 1:** Energy diagram for the electron-nuclear spin set in the lower right. Optical illumination (red lightning) initializes spin $S$ into $|m_S = 0\rangle$. The magnetic field is set at a value $B_m^{(\alpha)}$ so that $|\uparrow, 0, +1/2, \downarrow\rangle$ and $|\downarrow, -1, -1/2, \uparrow\rangle$ are degenerate.

We organize this article in the following way: In Section II, we introduce non-reciprocal nuclear spin dynamics via a reduced spin set comprising two separate nuclear spins whose mutual coupling is made possible by two interacting electron spins, each hyperfine-coupled to either nuclear spin. We generalize these ideas in Section III, to show how optical pumping and spin-lattice relaxation combine to produce a steady state gradient of nuclear polarization across the spin chain. Section IV focuses on ring-like structures to expose the formation of persistent nuclear spin currents driven by light. Sec. V discusses the impact of disorder in the spin chain and shows that strategies can be put in place to mitigate its effects. Finally, Sec. VI uses a model non-Hermitian Hamiltonian to associate the system dynamics with the presence of an exceptional point.

## II. NON-RECIPROCAL NUCLEAR SPIN TRANSPORT IN A FOUR-SPIN SET

The spin set in Fig. 1 provides a suitable framework to introduce some of the main ideas: We consider two nuclear spins $I_j = 1/2$, $j = 1,2$ each of them hyperfine-coupled to electron spins $S = 1$ and $S' = 1/2$, themselves interacting via a dipolar coupling $\mathcal{J}_d$. For simplicity, we assume both electronic spins feature identical, isotropic g-tensors, and that spin $S$ experiences a cylindrically symmetric crystal field. This system can be viewed as an idealized model of a molecule featuring a bi-radical, engineered so that one of the electronic spins — in the present case, spin $S$ — dynamically polarizes under optical excitation at a rate $\Gamma_{op}$. Related systems are presently being explored for improved forms of DNP[17-20], or as molecular spin qubits[21-27].

In the absence of illumination, we write the system Hamiltonian as

$$H^{(1,2)} = H_n + H_e + H_c + H_h + H_d, \quad (1)$$

where $H_n = -\omega_n I_1^z - \omega_n I_2^z$ and $H_e = \omega_e S^z + \omega_e S'^z$ respectively denote the nuclear and electronic Zeeman contributions, $H_c = D(S^z)^2$ is the crystal field on spin $S$, $H_h = A_{zz_1} S^z I_1^z + A_{zx_1} S^z I_1^x + A'_{zz_2} S'^z I_2^z + A'_{zx_2} S'^z I_2^x$ contains all secular hyperfine contributions, and $H_d = \frac{\mathcal{J}_d}{2}(S^+ S'^+ + S^- S'^-)$ is the secular dipolar coupling between the two electronic spins. In the above expressions, $I_j$ ($j = 1,2$) denotes the nuclear $^{13}$C spin operator, $S$ and $S'$ are the electronic spin operators, and $\omega_n = \gamma_n B$ and $\omega_e = |\gamma_e|B$ are the nuclear and electron Zeeman frequencies in the presence of a magnetic field $B$. Coefficients $A_{zz_1}$ and $A_{zx_1}$ ($A'_{zz_2}$ and $A'_{zx_2}$) denote the hyperfine tensor components coupling spins $I_1$ ($I_2$) and $S$ ($S'$), $\mathcal{J}_d$ is the dipolar coupling constant between $S$ and $S'$, and $D$ is the zero-field splitting of spin $S$, here assumed much greater than any hyperfine coupling constant.

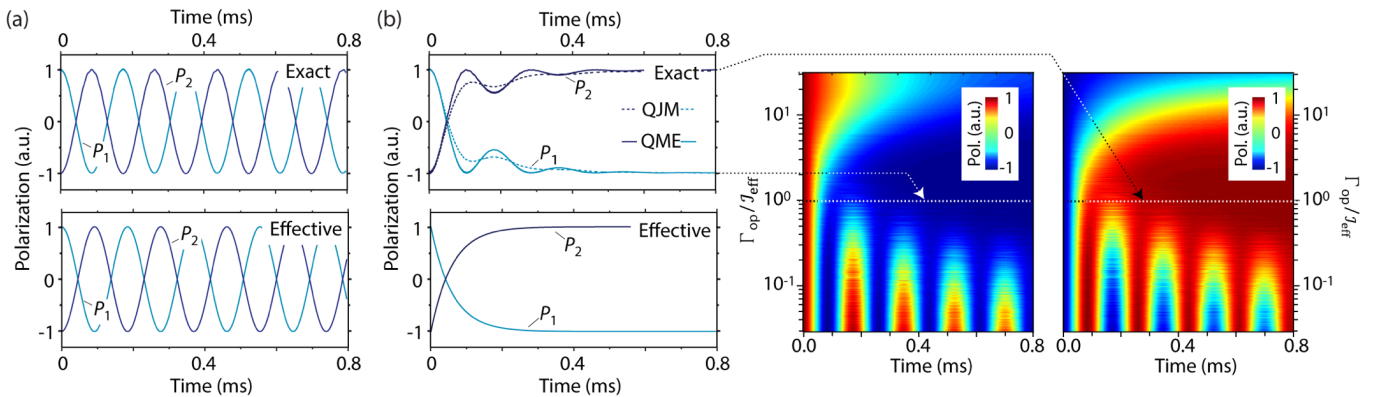

**Fig. 2:** One-directional transport of spin order. (a) Nuclear spin polarization $P_j$, $j = 1,2$ as a function of time for the initial state $|\uparrow, 0, +1/2, \downarrow\rangle$ (using notation $|m_{I_1}, m_S, m_{S'}, m_{I_2}\rangle$) as determined from an exact solution of the system Hamiltonian in Eq. (1) (top) or the effective nuclear spin Hamiltonian in Eq. (2) (bottom) in the absence of optical illumination. (b) (Left) Same as above but assuming optical spin pumping of $S$ at a rate $\Gamma_{op} \approx \mathcal{J}_{eff} = 17$ kHz. In the top plot, the solid and dashed lines respectively indicate a numerical calculation using the quantum master equation (QME) or a quantum jump Monte Carlo (QJM). (Right) Generalization for variable optical spin pumping rate $\Gamma_{op}$ obtained via QJM. In all calculations, we use $\mathcal{J}_d = 247$ kHz, $A_{zz_1} = A_{zx_1} = 13$ MHz, and $A'_{zz_2} = A'_{zx_2} = 4$ MHz, and we ignore $S'$ spin-lattice relaxation.

Unless explicitly noted, we set the magnetic field at (or near) $B_m^{(\alpha)}$ chosen so that states featuring opposite nuclear spin projections become degenerate* (see energy diagram in Fig. 1). Correspondingly, these two states hybridize and, despite the hyperfine coupling mismatch, reversible inter-nuclear polarization flow takes place in the dark thanks to the mediating action of the electron spins[28,29] (upper plot in Fig. 2a). We have already shown in prior work[30] that the above dynamics can be cast in the form of a simpler, nuclear-spin-only effective Hamiltonian

$$H_{\text{eff}}^{(1,2)} = \delta_{\text{eff}} I_1^z - \delta_{\text{eff}} I_2^z + \mathcal{J}_{\text{eff}}(I_1^+ I_2^- + I_1^- I_2^+), \quad (2)$$

an expression valid for variable, small magnetic field shifts $\delta_{\text{eff}} = 2\left|B - B_m^{(\alpha)}\right| |\gamma_e|$ near $B_m^{(\alpha)}$. The effective inter-nuclear coupling in Eq. (2) is given by[30]

$$\mathcal{J}_{\text{eff}} = \frac{\mathcal{J}_d}{2} \sin(\theta_1/2) \sin[(\theta_2^- - \theta_2^+)/2]. \quad (3)$$

where $\tan(\theta_1) = A_{zx_1}/(A_{zz_1} + \omega_I)$, $\tan(\theta_2^{m_{S'}}) = m_{S'} A'_{zx_2}/(m_{S'} A'_{zz_2} - \omega_I)$, and $m_{S'} = \pm 1/2$ are the projections of spin $S'$. Conceptually, $H_{\text{eff}}$ expresses inter-electronic cross-relaxation as a process of nuclear spin transport; as shown in the lower panel of Fig. 2a, the model reproduces the exact numerical solution reasonably well. We emphasize the symmetry in the Hamiltonian ensures there is no preferential bias in the transport of nuclear polarization, i.e., polarization flows reversibly between the two nuclear spins in the chain.

To quantitatively assess the impact of spin-lattice relaxation and electron spin pumping we formulate the system dynamics as a quantum master equation[31] (QME), namely,

$$\dot{\rho} = -i[H^{(1,2)}, \rho] + \mathcal{L}^{(1,2)}(\rho), \quad (4)$$

where $\rho$ denotes for the 4-spin density matrix, and $\mathcal{L}^{(1,2)}(\cdot)$ is the GKS-Lindblad generator. Ignoring for now spin-lattice relaxation, we write

$$\mathcal{L}^{(1,2)}(\rho) = C_{\text{op}} \rho C_{\text{op}}^\dagger - \frac{1}{2}\{C_{\text{op}}^\dagger C_{\text{op}}, \rho\}, \quad (5)$$

where the optical pumping operator $C_{\text{op}}$ acts on spin $S$ as

$$C_{\text{op}} = \sqrt{\Gamma_{\text{op}}}[|0\rangle\langle -1| + |0\rangle\langle +1|], \quad (6)$$

and $\{\cdot\}$ denotes anti-commutation.

Fig. 2b shows the results for the case $\Gamma_{\text{op}} \approx \mathcal{J}_{\text{eff}}$: Unlike the case in Fig. 2a, polarization flows from $I_1$ to $I_2$ but the converse process is quenched. Qualitatively, this one-directional response stems from the chosen optical pumping rate, polarizing spin $S$ after half an evolution cycle (i.e., after positive spin polarization passes from $I_1$ to $I_2$), and hence driving the system away from degeneracy, towards a state where no further nuclear spin flow occurs. As shown in the right panels of Fig. 2b, this rate amounts to a 'critical damping' of the transfer dynamics: Bi-directional polarization transfer reappears when $\Gamma_{\text{op}}$ is sufficiently low, whereas strong spin pumping gradually quenches spin transport altogether (as the laser pins spin $S$ into $m_S = 0$). Reassuringly, similar results emerge from a quantum-jump Monte Carlo[32,33] (QJM), an arguably more intuitive method where we average over multiple trajectories concatenating unitary evolutions and random projections of spin $S$ into $m_S = 0$ with unit time probability $\Gamma_{\text{op}}$ (dashed traces in the upper left panel of Fig. 2b).

For future reference, here too one can quantitatively model the system's non-Hermitian dynamics via a nuclear-spin-only master equation governed by a one-directional coupling transferring nuclear population from state $|\uparrow\downarrow\rangle$ to $|\downarrow\uparrow\rangle$ but leaving all other populations unchanged. We thus write

$$\dot{\rho}_n = \mathcal{L}_{\text{eff}}^{(1,2)}(\rho_n), \quad (7)$$

where

$$\mathcal{L}_{\text{eff}}^{(1,2)}(\rho_n) = C_{\text{eff}} \rho_n C_{\text{eff}}^\dagger - \frac{1}{2}\{C_{\text{eff}}^\dagger C_{\text{eff}}, \rho_n\} \quad (8)$$

is the Lindblad operator of the nuclear spin pair with $C_{\text{eff}} = \sqrt{\mathcal{J}_{\text{eff}}} I_1^- I_2^+$ (see lower panel in Fig. 2b). In the above expressions, $I_j^\pm$, $j = 1,2$ are the nuclear raising/lowering operators, and $\rho_n$ is the reduced density matrix representing the state of the nuclear spins. The built-in non-reciprocity of the set is implicit in the definition of the collapse operator $C_{\text{eff}}$, selectively moving positive polarization from nuclear spin 1 to 2 (i.e., the reverse process is forbidden).

### III. INTERPLAY BETWEEN GAIN AND LOSS IN LINEAR SPIN CHAINS

To investigate how these ideas play out in more complex spin sets, we introduce the notion of a 'dressed' nucleus, whereby spin $I$ interacts with adjacent nuclei through non-equivalent magnetic 'bonds' (Fig. 3a). For presentation purposes, we start by considering the linear, open-ended chain shown in Fig. 3b, comprising a collection of dressed nuclei whose spin dynamics we break down into a series of consecutive stages, (*i*) through (*v*). The latter allows us to express nuclear transport as a stepped process involving coherent effective couplings between nuclei as well as electronic spin-lattice relaxation and spin pumping (with the understanding, however, that all of the above acts simultaneously in a real system). Further, we assume a well-defined initial spin configuration where the nuclear (electron) spin at site $j - 2$ $(j - 1)$ is 'up' while all others are 'down'. Under these conditions, the spin set formed by dressed nuclei $j - 2$ and $j - 1$ effectively reproduces the dynamics already observed in Fig. 2, namely, nuclear magnetization moves one site to the right because the converse process is hindered by the optical pumping of spin $S$ (stages (*i*) through (*iii*) in Fig. 3b). In this case, the amplitude of the effective coupling — slightly different from that derived above for the 4-spin model — is given by

$$\mathcal{J}_{\text{eff}} = \frac{\mathcal{J}_d}{2} \sin\left(\frac{\theta_0^- - \theta_{-1}^-}{2}\right) \sin\left(\frac{\theta_0^- - \theta_0^+}{2}\right), \quad (9)$$

where $\tan \theta_{m_S}^{m_{S'}} = \frac{(m_S A_{zx} + m_{S'} A'_{zx})}{(m_S A_{zz} + m_{S'} A'_{zz} - \omega_n)}$, and $m_S = 0, -1$ ($m_{S'} = \pm 1/2$) are the projections of spin $S$ (spin $S'$). Note that we drop the site index in the hyperfine components by relying on the assumption of translational invariance (see Appendix A for details). Further directional transport is ensured by electron spin-lattice relaxation, as it keeps the system from being trapped in configurations where cross-relaxation is disallowed (stages (*iv*) and (*v*)); by contrast, nuclear relaxation — typically much slower — is unnecessary and can be neglected. Qualitatively, the underlying dynamics can be loosely pictured as a stochastic sequence of directional hopping events, here enabled by the simultaneous presence of dynamic polarization and spin-lattice relaxation (Fig. 3b). We stress that the optical spin pumping simultaneously acts on every spin $S$ of the chain, i.e. spin transport takes place without assuming any local manipulation. Further, despite considering a specific electron/nuclear spin state when describing Fig. 3b, such an assumption is unnecessary and is made only to clarify the mechanism underlying the one-directional spin transport dynamics.

---

* We use the supra-index $\alpha$ to distinguish this field from $B_m^{(\beta)}$ where states $|\downarrow, 0, +1/2, \uparrow\rangle$ and $|\uparrow, -1, -1/2, \downarrow\rangle$ become degenerate and analogous — though not identical — nuclear spin dynamics occurs[28].

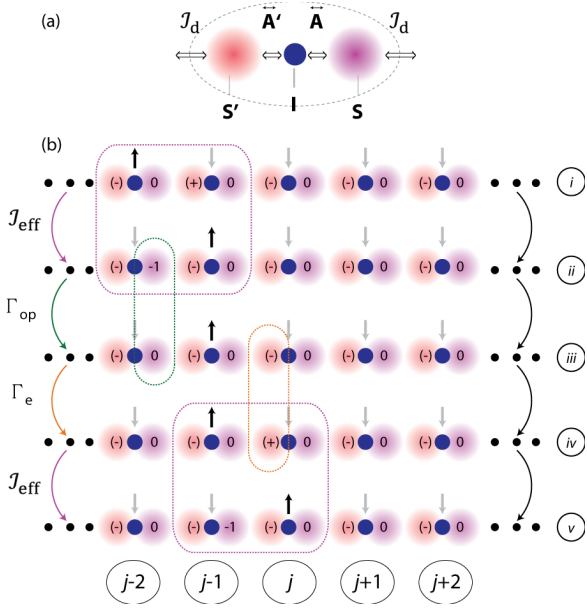

**Fig. 3:** One-directional nuclear spin transport through gain and loss. (a) Schematics of a 'dressed' nuclear spin. (b) In the simultaneous presence of optical excitation and spin relaxation, positive (negative) nuclear spin polarization propagates to the right (left). In panel (b), plus and minus signs (up and down arrows) indicate the projections of electrons (nuclei) with spin number ½.

The upper panels in Figs. 4a through 4c show the averaged time evolution of the nuclear spin polarization for a chain with $N = 4$ sites as determined from a QJM. Unlike the results in Fig. 2, we assume that electron spins are subject to spin-lattice relaxation with a rate $\Gamma_e < \Gamma_{\text{op}}$, and they are all initially in a maximally mixed state (i.e. unpolarized). For the nuclear spins, we consider alternative initial spin configurations, including the case where all nuclear spins $j = 1 \ldots N$ in the chain are initially unpolarized (i.e., $\rho_n^{(j)} = \mathbb{I}$). Remarkably, the system evolves in all cases towards a state where positive (negative) nuclear polarization concentrates at the right (left) end of the chain, even if the starting nuclear polarization is null. Unlike in DNP, the net nuclear magnetization of the chain remains unchanged, implying that electronic gain (i.e., optical spin pumping) and loss (i.e., spin-lattice relaxation) combine herein to redistribute — not generate — nuclear polarization.

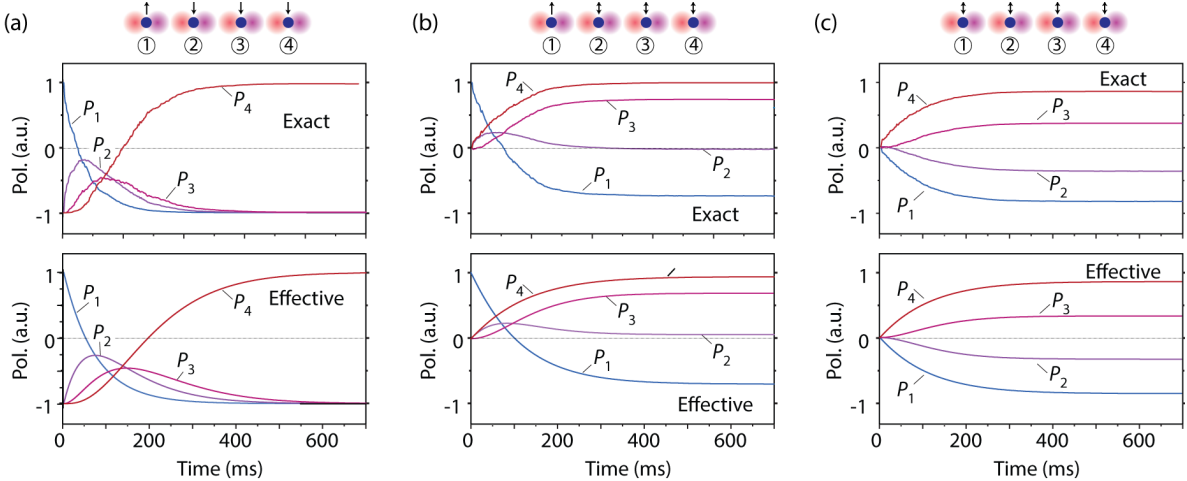

**Fig. 4**: Non-reciprocal spin dynamics in multi-site chains. (a) Nuclear spin polarization as a function of time for a four-site chain as determined from QJM (top) or the effective nuclear spin Lindbladian (bottom). Arrows in the upper cartoon indicate the initial nuclear spin polarization, positive for the nuclear spin $j = 1$, negative for all others. (b) Same as before but assuming nuclear spins in sites $j = 2 \cdots 4$ are unpolarized (two sided arrows in the cartoon). (c) Same as in (a) but for a case where all nuclear spin are initially unpolarized. In all instances, we assume all electronic spins are initially unpolarized and use $\mathcal{J}_d = 62$ kHz, $A_{zz} = A_{zx} = 13$ MHz, and $A'_{zz} = A'_{zx} = 4$ MHz, which corresponds to $\mathcal{J}_{\text{eff}} = 0.547$ kHz. As before, we set $\Gamma_{\text{op}} \approx \mathcal{J}_{\text{eff}}$ but assume electronic spins relax with a rate $\Gamma_e = 0.1$ kHz. For the QJM results, the number of averages (6 in this case) is chosen to minimize the statistical error. All other conditions as in Fig. 2.

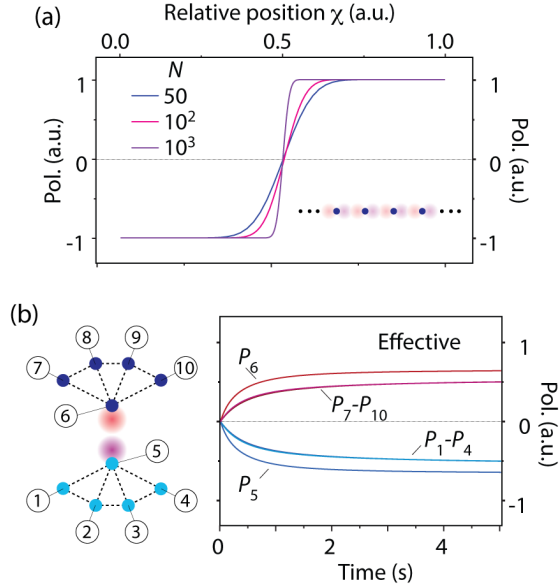

**Fig. 5**: (a) Calculated steady-state nuclear spin polarization as a function of the relative position $\chi = j/N$ where $j = 1 \ldots N$ is the site index along the chain. (b) Nuclear spin polarization as a function of time for the 'tree-like' spin structure on the left. Dashed lines indicate bi-directional inter-nuclear couplings with value $\mathcal{J}_b = 547$ Hz; the effective directional coupling between nuclei 5 and 6 is given by a collapse operator $C_{\text{eff}} = \sqrt{\Gamma_{\text{eff}}} I_5^- I_6^+$, where $\Gamma_{\text{eff}}$ is 13 Hz.

Fortunately, the Lindblad formalism introduced in Eq. (7) for the nuclear pair can be easily generalized to multi-site chains, namely, we write

$$\dot{\rho}_n = \mathcal{L}_{\text{eff}}^{(1,2,\ldots,N)}(\rho_n), \quad (10)$$

where

$$\mathcal{L}_{\text{eff}}^{(1,2,\ldots,N)}(\rho_n) = \sum_{j=1}^{N-1} \left[ C_{\text{eff},j} \rho_n C_{\text{eff},j}^\dagger - \frac{1}{2} \{ C_{\text{eff},j}^\dagger C_{\text{eff},j}, \rho_n \} \right] \quad (11)$$

and $C_{\text{eff},j} = \sqrt{\Gamma_{\text{eff}}} \, I_j^- I_{j+1}^+$ with the exact value of $\Gamma_{\text{eff}}$ adjusted to take into account the combined effects of spin pumping and electron spin relaxation (see Appendix B). As a comparison between the upper and lower panels in Figs. 4a-4c shows, we attain good agreement with the QJM calculations in all cases, regardless the initial nuclear spin configuration.

The ability to model the system dynamics through the simplified framework of Eqs. (10) and (11) allows us to extend our theoretical analysis to large nuclear spin sets, beyond the capabilities of present numerical methods. Fig. 5a shows an example where we plot the steady state nuclear polarization as a function of the fractional position in the chain $\chi = j/N$, $j = 1 \ldots N$ for a variable number of sites $N$ assuming an initial state with all spins unpolarized. The same trends observed before generalize to yield a limit density matrix

$$\rho_n(t \to \infty) = \frac{1}{2^N} \sum_{i=0}^{N} \binom{N}{i} [\otimes_{j=1}^{i} |\downarrow\rangle\langle\downarrow|] \otimes [\otimes_{j=i+1}^{N} |\uparrow\rangle\langle\uparrow|], \quad (12)$$

with $|\uparrow\rangle$, $|\downarrow\rangle$ representing the two possible nuclear spin projections at each site. This expression follows from the maximally mixed initial nuclear spin state: Every possible nuclear configuration has the same statistical weight and it evolves towards an ordered 'domain wall state', where all spins at the right are positively polarized and spins to the left are negatively polarized. For example, in a four-spin chain, states $|\downarrow\uparrow\uparrow\uparrow\rangle$, $|\uparrow\downarrow\uparrow\uparrow\rangle$, $|\uparrow\uparrow\downarrow\uparrow\rangle$, and $|\uparrow\uparrow\uparrow\downarrow\rangle$ converge to $|\downarrow\uparrow\uparrow\uparrow\rangle$. In addition, completely ordered states do not evolve at all, e.g. $|\downarrow\downarrow\downarrow\downarrow\rangle$. Equation (12) then represents the asymptotic nuclear state by counting the contributions to each domain-wall state.

Lastly, we note that similarly biased transport dynamics can be recreated in spin sets other than a linear chain. Fig. 5b shows one illustration where we numerically calculate the dynamics of a tree-like nuclear spin molecule engineered such that only the effective coupling along the 'trunk' (i.e., the central bond) is one-directional; interestingly, we find that nuclei in the 'tree top' and at its 'roots' polarize uniformly but with opposite signs.

## IV. PERSISTENT NUCLEAR SPIN CURRENTS IN RING-LIKE ARRAYS

While the discussion thus far centered on open-ended arrays, periodic spin chains give us the opportunity to uncover a distinct response. This is shown in Figs. 6a and 6b, where we consider a three-site electron/nuclear ring, and calculate the time evolution of the site-selective nuclear polarization for two different initial conditions (upper and lower panels in Fig. 6b). Irrespective of whether we resort to the nuclear Lindblad model or conduct a full QJM (solid and faint traces, respectively), we find the system evolves into a state of uniform polarization whose amplitude correlates with the net initial magnetization, negative in the first case, null in the other.

We emphasize that although the local nuclear polarization in rings and linear arrays behaves differently (forming opposite domains in one case, spreading uniformly in the other), the underlying gain/loss, dissipative dynamics (one-directional in both instances) must remain unchanged. To expose the biased nature of the transport at play, we turn our attention to the nuclear spin current $K_{j,j+1}$ between sites $j$ and $j+1$, which can be connected to the on-site nuclear polarization via the continuity equation[34-36]

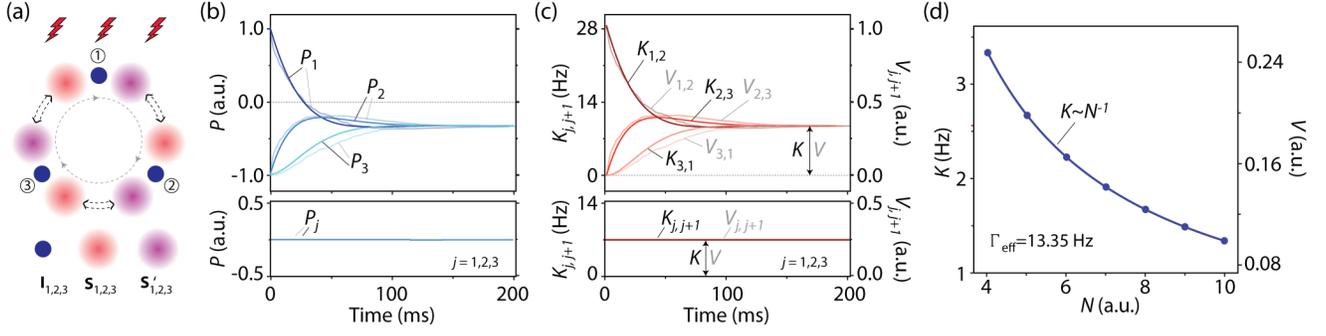

**Fig. 6:** Persistent spin current in periodic chains. (a) Schematic of a three-unit electron/nuclear spin ring. (b) Time evolution of the nuclear spin polarization under optical excitation starting from an initial nuclear spin state $|\uparrow,\downarrow,\downarrow\rangle$ (top) or from a state where all nuclear spins are unpolarized (bottom). Solid (faint) traces indicate evolution as determined from the effective nuclear spin Lindbladian (QJM of the full spin Hamiltonian). (c) Time evolution of the nuclear spin current $K_{j,j+1}$ and voltage $V_{j,j+1}$ between sites $j$ and $j+1$ using the effective nuclear Lindbladian (solid traces) or the full Hamiltonian under QJM dynamics (faint traces); the initial state is $|\uparrow,\downarrow,\downarrow\rangle$ (top) or a state of unpolarized nuclear spins (bottom). Even in the absence of nuclear polarization, a non-zero spin current is seen to persist at long times. (d) Persistent spin current $K$ as a function of the number of nuclei $N$ in the ring assuming an initial state of the form $|\uparrow,\downarrow,\downarrow\cdots\downarrow\rangle$. Throughout these calculations the number of averaged QJM trajectories is 28; all other conditions as in Fig. 4.

$$\frac{d}{dt}I_j^z = K_{j-1,j} - K_{j,j+1}. \quad (13)$$

While a numerical calculation that takes into account the full electron/nuclear spin set can be implemented via QJM, it is also possible to evaluate the spin current via a nuclear-spin-only Lindblad formulation in the Heisenberg representation. In particular, the left-hand side of Eq. (13) can be cast as

$$\frac{d}{dt}I_j^z = \mathbb{L}_{\text{eff}}^{(1,2,\ldots,N)}(I_j^z) =$$
$$= \sum_{i=1}^{N-1}\left[C_{\text{eff},i}^{\dagger}I_j^z C_{\text{eff},i} - \frac{1}{2}\{C_{\text{eff},i}^{\dagger}C_{\text{eff},i}, I_j^z\}\right], \quad (14)$$

where $\mathbb{L}_{\text{eff}}^{(1,2,\ldots,N)}$ is the Hilbert-Schmidt adjoint of $\mathcal{L}_{\text{eff}}^{(1,2,\ldots,N)}$. Note that since $C_{\text{eff},i}$ selectively connects spins $i$ and $i+1$, only two terms in the above generator $\mathbb{L}_{\text{eff}}^{(1,2,\ldots,N)}(\cdot)$ remain after summation, each one coupling the $j$-th nuclear spin to its left and right neighbors, i.e.,

$$\frac{d}{dt}I_j^z = \sum_{i=j-1}^{j}\left[C_{\text{eff},i}^{\dagger}I_j^z C_{\text{eff},i} - \frac{1}{2}\{C_{\text{eff},i}^{\dagger}C_{\text{eff},i}, I_j^z\}\right]. \quad (15)$$

A comparison between Eqs. (13) and (15) hence suggests the natural definition

$$K_{j,j+1} \equiv -C_{\text{eff},j}^{\dagger}I_j^z C_{\text{eff},j} + \frac{1}{2}\{C_{\text{eff},j}^{\dagger}C_{\text{eff},j}, I_j^z\}, \quad (16)$$

which, after some algebra yields

$$K_{j,j+1} = \Gamma_{\text{eff}}I_j^+I_j^-I_{j+1}^-I_{j+1}^+ = \Gamma_{\text{eff}}|\uparrow_j\downarrow_{j+1}\rangle\langle\uparrow_j\downarrow_{j+1}|. \quad (17)$$

Fig. 6c shows the results assuming, as before, two alternative initial conditions. Consistent with the QJM results for the whole electron/nuclear chain (faint traces), the system stabilizes after a variable, state-dependent transient to ultimately attain a uniform ring current, persisting even in the absence of net nuclear magnetization.

Equation (17) is reminiscent of Ohm's law, namely, spin current emerges as a consequence of a local 'potential difference' $V_{j,j+1} \equiv |\uparrow_j\downarrow_{j+1}\rangle\langle\uparrow_j\downarrow_{j+1}|$ that depends on the probability of finding opposite polarizations in adjacent nuclear sites. The unit-length 'conductivity' $\Gamma_{\text{eff}} \propto \Gamma_e$ — governing the spin current in the limit $\Gamma_{\text{op}} > \Gamma_e$ used herein, see Appendix B — shows that spin-lattice relaxation is a central component of spin transport in these systems. We reinforce this picture in Fig. 6d where we plot the persistent spin current $K$ (local voltage $V$) as a function of the number $N$ of nuclear sites in the ring. We find that $K$ changes inversely with $N$, as expected for a wire whose 'resistance' grows linearly with its length.

For completeness, we mention that a similar calculation in linear chains yields null persistent currents preceded by non-zero transients, consistent with the open-ended structure of these arrays (Fig. 7a). Even though Eq. (17) breaks down (in general) due to the lack of periodicity, the buildup of a non-zero nuclear polarization gradient shown in Fig. 7b can be viewed as the result of a 'magnetic charge' separation process. Such a non-uniform distribution leads to a persistent voltage difference, corresponding to the operator $|\uparrow_N\downarrow_1\rangle\langle\uparrow_N\downarrow_1|$ — with no net current — between the two extremes of the spin wire as shown in Fig. 7c.

## V. NUCLEAR SPIN TRANSPORT IN IMPERFECT CHAINS

Consistent with the resistive character of the transport, we find that the dynamics is sensitive to the presence of 'defects' (i.e., imperfections) in the chain. This is shown in Fig. 8, where we consider the case of a three-site linear chain in which the dressed nucleus at its center is different from the other two; the latter can be attained, e.g., by modifying the hyperfine coupling to spin $S'$ (see schematics in Fig. 8a). Even in the presence of electron spin pumping, the energy mismatch inhibits multi-site polarization exchange and hence the nuclear spin set develops no collective order, regardless the applied magnetic field (main plot in Fig. 8a). Note that the 'cross points' $B_{\text{m}}^{(\alpha)} + \delta B_{\text{m}}^{(\alpha_{j,j+1})}$ — where adjacent nuclei $j, j+1$ polarize in opposite directions — are simply a fortuitous consequence of local matching between the nuclear and electronic energy differences, and thus cannot induce long-range transport along the chain.

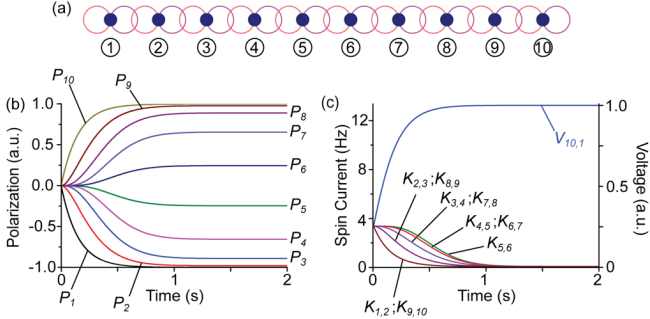

**Fig. 7:** (a) Ten-site spin chain. Empty circles indicate effective inter-nuclear couplings. (b) Time dependence of the nuclear spin polarization in the above chain as determined from the effective nuclear spin Lindbladian. (c) Nuclear spin current between different consecutive sites in the chain as a function of time. The voltage $V_{10,1}$ is also evaluated. We assume that the initial state is fully unpolarized and use $\Gamma_{\text{eff}} = \left|\ln\left(\frac{7}{8}\right)\right|\Gamma_{\text{e}}$, with $\Gamma_{\text{e}} = 0.1$ kHz.

We regain the spin dynamics observed in Fig. 4 with the aid of the protocol in Fig. 8b, articulating simultaneous electron spin pumping and magnetic field modulation. Motivating this scheme is the intuition that one can move polarization pair-wise, simply by bringing the magnetic field to the value required to enable a local 'flip-flop' between adjacent sites $j-1$ and $j$, then changing it to couple sites $j$ and $j+1$. Correspondingly, successive field sweeps over a suitably broad range must gradually move positive (negative) polarization to the right (left) of the chain, thus recreating the prior spin dynamics. Note that the symmetric nature of the field modulation — featuring equally fast low-to-high and high-to-low ramps — ensures no net polarization emerges from Landau-Zener dynamics at level anti-crossings[37,38].

Fig. 8c displays the results for a variable field modulation frequency $\Gamma_B$ assuming the field sweeps the range in Fig. 8a. The protocol recreates the spin dynamics already observed in Fig. 4c although the process is comparatively slower, a consequence of the finite time required to complete a field sweep. Increasing $\Gamma_B$ can only mitigate this limitation partially (right panels in Fig. 8c) because polarization transfer is gradually inhibited if the effective time during which pair-wise energy matching is attained becomes shorter than the inverse effective coupling between adjacent nuclei. Lastly, we mention that a similar response is observed in related defects where the hyperfine coupling to spin $S$ (or to both spin $S$ and $S'$) are changed in one (or more) sites along the chain. Nonetheless, we observe the gradual appearance of net polarization once disorder in the chain exceeds some critical threshold. Other types of defects are insensitive to the correction protocol in Fig. 8b, and can lead to different spin

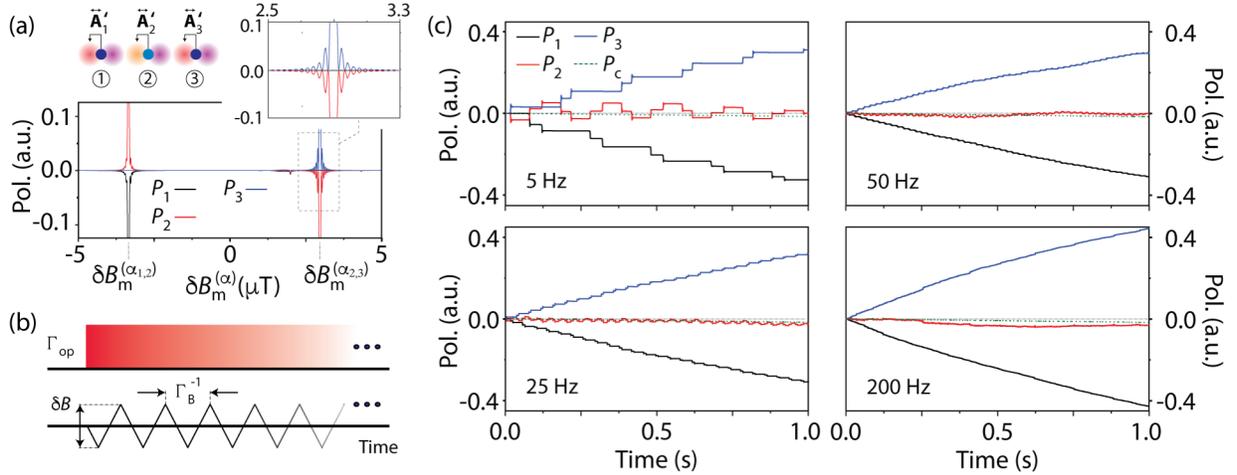

**Fig. 8:** Spin dynamics of a defective dressed nuclear spin chain. (a) Three-site chain with a defect formed by a dressed nucleus in site 2 featuring a hyperfine tensor $\vec{A}'_1 = \vec{A}'_3 = 0.885\, \vec{A}'_2$. Under optical excitation and a static magnetic field, the nuclear polarization at all sites is null, except at 'cross points' $B_{\text{m}}^{(\alpha)} + \delta B_{\text{m}}^{(\alpha_{j,j+1})}$ where adjacent nuclei $j, j+1$ polarize in opposite directions. (b) To re-establish one-directional polarization transport, we modulate the magnetic field around $B_{\text{m}}^{(\alpha)}$ with amplitude $\delta B$ and frequency $\Gamma_B$ during electronic spin pumping. (c) Nuclear spin polarization as a function of time obtained after a single QJM run for the full electron/nuclear spin set. In each plot, we assume a different field modulation frequency $\Gamma_B$ (lower corner in each plot) and a field sweep amplitude $\delta B_{\text{m}}^{(\alpha)} = 10$ μT centred at $B_{\text{m}}$. In these simulations, we use $A'_{zz_1} = A'_{zz_3} = A'_{zx_1} = A'_{zx_3} = 3.75$ MHz and $A'_{zz_2} = A'_{zx_2} = 4.25$ MHz; all other conditions as in Fig. 2. In panels (a) and (c), $P_j$ denotes the nuclear polarization at site $j = 1 \ldots 3$, and $P_c$ indicates the net nuclear polarization of the chain.

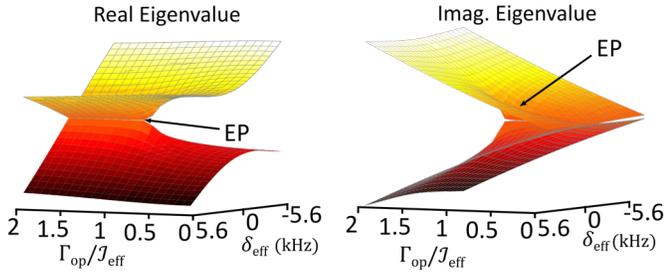

**Fig. 9:** Riemann surface of the eigenvalues of the non-Hermitian Hamiltonian $H_{nH}^{(1,2)}$ versus parameters $\delta_{\text{eff}}, \Gamma_{\text{op}}$. An exceptional point appears at $\delta_{\text{eff}} = 0$ and $\Gamma_{\text{op}} = \mathcal{J}_{\text{eff}}$. Throughout these calculations the coupling between electron spins is $\mathcal{J}_d = 62$ kHz; all other conditions as in Fig. 2.

dynamics (such as the formation of multiple 'domains', not shown here for brevity).

## VI. TOWARDS A NON-HERMITIAN HAMILTONIAN

Our approach to the dynamics of gain and loss so far has been driven by an explicit evaluation of the system dynamics using the Lindblad formulation, both in the complete electron-nuclear and purely nuclear scenarios. An alternative strategy would be to analyze the same physics replacing Linbladians by some appropriate non-Hermitian Hamiltonian (nHH). While a rigourous derivation goes beyond the scope of this article, we can straightforwardly reshape the effective dynamics discussed in Section II by proposing a nHH in the role of the optical pumping of spin $S$, $H_{op} = i\Gamma_{\text{op}}|0\rangle\langle 0| - i\Gamma_{\text{op}}|-1\rangle\langle -1|$. Then, the complete nHH for the four spin system in Fig. 1 is $H_{nH}^{(1,2)} = H^{(1,2)} + H_{op}$, with $H^{(1,2)}$ as defined in Eq. (1). In the subspace spanned by $|\uparrow, 0, +1/2, \downarrow\rangle$ and $|\downarrow, -1, -1/2, \uparrow\rangle$ (using notation $|m_{I_1}, m_S, m_{S'}, m_{I_2}\rangle$) we get

$$H_{nH}^{(1,2)} = \begin{pmatrix} \delta_{\text{eff}} + i\Gamma_{\text{op}} & \mathcal{J}_{\text{eff}} \\ \mathcal{J}_{\text{eff}} & -\delta_{\text{eff}} - i\Gamma_{\text{op}} \end{pmatrix} \quad (18)$$

where $\delta_{\text{eff}}$ plays the role of a detuning parameter, vanishing at the matching condition $B = B_m^{(\alpha)}$.

Figure 9 shows the Riemann surfaces of the complex eigenvalues of $H_{nH}^{(1,2)}$ versus parameters $\delta_{\text{eff}}, \Gamma_{\text{op}}$. An exceptional point arises when $\delta_{\text{eff}} = 0$ and $\Gamma_{\text{op}} = \mathcal{J}_{\text{eff}}$. This last condition, as discussed before, is precisely our choice for optimal one-directional spin flips.

## VII. CONCLUSIONS

We studied the dynamics of nuclear spin chains whose interactions are mediated by pairs of coupled electron spins of different types, and showed that the otherwise reciprocal nuclear polarization transport becomes one-directional upon externally pumping one of the electron spin species. Ultimately reflecting the asymmetric nature of the effective inter-nuclear bond, this process requires a suitable combination of electron spin polarization gain and loss, i.e., non-reciprocity vanishes (in general) if only spin-lattice relaxation or spin pumping are active. We attain optimal transport when the optical pumping rate and the effective coupling constant are equal and greater than the electron spin-lattice relaxation rate, i.e., $\mathcal{J}_{\text{eff}} \approx \Gamma_{\text{op}} > \Gamma_e$; in this regime, a net transport of nuclear polarization takes place with rate $\Gamma_{\text{eff}} \propto \Gamma_e$.

Although the same underlying transport mechanisms are at play, the steady state of open and closed spin chains are markedly different: Net nuclear spin magnetization of opposite signs builds up on either end of a linear chain, while ring-like arrays feature a uniform distribution of polarization. Importantly, however, no polarization is created nor destroyed in either case, implying that, although related, this process differs from dynamic nuclear polarization. Diffusive in nature, nuclear spin transport in these chains can be sensitive to disorder, though we have shown that the impact of some imperfections (such as heterogeneous effective couplings between nuclei) can be mitigated through the use of magnetic field sweeps.

While a rigorous comparison between the present results and prior work in other non-Hermitian systems falls beyond the scope of this manuscript, we find some interesting analogies and differences. For example, similar to ring-like arrays of opto-mechanical resonators connected to a thermal bath[39], here we observe persistent one-directional transport, though in our case the system displays no long-range coherences and separately driving individual chain sites is unnecessary. By the same token, the 'spontaneous' formation of positive and negative polarization domains in the linear spin chains considered herein is somewhat complementary to the non-Hermitian 'skin effect'[40-43] where, regardless the initial conditions, modes localize at the interface between sections of the array with different topologies. As already observed in other physical platforms — including a mechanical metamaterial[44], a topolectrical circuit[45], and in photonic arrays[6,46] — it should be possible to recreate equivalent spin-based dynamics by engineering chains featuring alternating dressed nuclei with anisotropic and isotropic effective bonds. Note that the non-trivial impact of topology on the steady state of the spin system makes the inverse design problem — namely, how to engineer relaxation given a target nuclear spin state — an intriguing research direction in its own right. Besides their intrinsic fundamental interest, this type of non-Hermitian spin set imbued with non-trivial topology could find application in metrology, for example, by implementing external drive protocols (i.e., laser excitation, magnetic field, etc.) that capitalize on the sensitivity of topological transitions to changes in the environment[47].

## APPENDIX A: DERIVATION OF $\mathcal{J}_{\text{eff}}$ FOR THE MULTI-SITE SPIN CHAIN

The Hamiltonian governing the dynamics in a 1-D array composed by $N$ dressed spins is given by

$$H^{(1,2,\dots,N)} = \sum_{j=1}^{N-1}\left[H_{u,j} + \frac{\mathcal{J}_d}{2}\left(S_j^+ S'^+_{j+1} + S_j^- S'^-_{j+1}\right)\right] + H_{u,N}, \quad (A.1)$$

where we define the 'unit cell' Hamiltonian as

$$H_u = -\omega_n I^z + \omega_e S^z + \omega_e S'^z + D(S^z)^2 + A_{zz}S^z I^z$$
$$+ A_{zx}S^z I^x + A'_{zz}S'^z I^z + A'_{zx}S'^z I^x. \quad (A.2)$$

To derive the effective coupling between nuclear spins we consider the set formed by two adjacent dressed nuclei (Fig. 10). This system is virtually identical to that considered in Section II except that the couplings with the electron spins at both ends of the chain must be taken into account. For linear arrays, we assume translational invariance, i.e. the hyperfine couplings are the same at each 'unit cell', so we drop the site index from each hyperfine component, but keep the prime label to indicate the coupling to $S'$ spins. Furthermore, we also assume that hyperfine interactions are stronger than the nuclear Zeeman couplings, so we do a partial diagonalization using the quantization axes for nuclear spins $I_1$ and $I_2$ defined by

$$\vec{Z}(m_S, m_{S'}) = (m_S A_{zx} + m_{S'} A'_{zx})\hat{x}$$
$$+ (m_S A_{zz} + m_{S'} A'_{zz} - \omega_n)\hat{z}, \quad (A.3)$$

where the laboratory frame has been chosen so that the $z$-axis coincides with the axis of the crystal field in spin $S$. The norms of these vectors are related to the strength of the hyperfine interactions, and will be used below,

$$\Delta^{m_{S'}}_{m_S} = |\vec{Z}(m_S, m_{S'})| =$$
$$\sqrt{(m_S A_{zx} + m_{S'} A'_{zx})^2 + (m_S A_{zz} + m_{S'} A'_{zz} - \omega_n)^2}. \quad (A.4)$$

The corresponding rotation angles required to transform into this eigen-frame representation are respectively defined by

$$\tan\theta^{m_{S'}}_{m_S} = \frac{(m_S A_{zx} + m_{S'} A'_{zx})}{(m_S A_{zz} + m_{S'} A'_{zz} - \omega_n)}, \quad (A.5)$$

Table A.1 contains the matrix representation of this 2-site dressed nuclear spin Hamiltonian in the subspace corresponding to $m_{S'_1} = \downarrow$, $m_{S_2} = 0$. These are considered 'dummy' electron spins located at both ends of the array. As before, we find that nuclear spin flip-flops are possible provided that the appropriate energy-matching condition is met. For instance, using the notation $|m_{S'_1}, m_{I_1}, m_{S_1}, m_{S'_2}, m_{I_2}, m_{S_2}\rangle$, the states $|-1/2, \uparrow, 0, +1/2, \downarrow, 0\rangle$ and $|-1/2, \downarrow, -1, -1/2, \uparrow, 0\rangle$ become degenerate if

$$-\Delta_0^+ - \Delta_0^- = D - 2\omega_e - \Delta_0^- + \Delta_{-1}^-, \quad (A.6)$$

which defines an equation for the 'matching' magnetic field $B = B_m^{(\alpha)}$. Note that the time scale for the nuclear flip-flop processes, $|\uparrow\downarrow\rangle \leftrightarrow |\downarrow\uparrow\rangle$, is defined by the rate

$$\langle -1/2, \downarrow, -1, -1/2, \uparrow, 0|H^{(1,2)}|-1/2, \uparrow, 0, +1/2, \downarrow, 0\rangle$$
$$\equiv \mathcal{J}_{eff} = \frac{\mathcal{J}_d}{2}\sin\left(\frac{\theta_0^- - \theta_{-1}^-}{2}\right)\sin\left(\frac{\theta_0^- - \theta_0^+}{2}\right). \quad (A.7)$$

An equivalent situation can be found for the condition

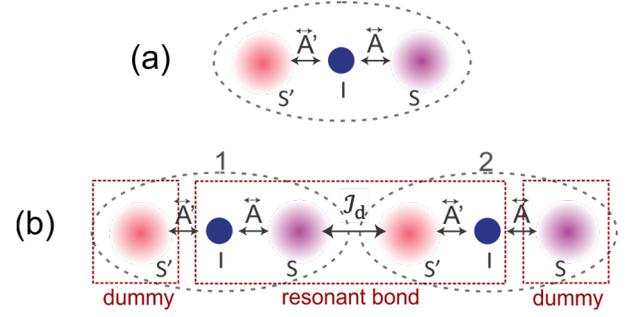

**Fig. 10.** (a) Electron-dressed nuclear spin comprising one nuclear spin $I$ and two electron spins, $S$ and $S'$. Spin $S$ has a spin number equal to 1, is optically polarizable, and is coupled to nuclear spin $I$ by the hyperfine tensor $\vec{A}$. Spin $S'$ is a spin-1/2 particle and is coupled to nuclear spin $I$ by the hyperfine tensor $\vec{A}'$. (b) Two dressed nuclear spins (at sites 1 and 2, dashed ovals) interacting via a dipolar coupling $\mathcal{J}_d$ between spins $S_1$ and $S'_2$.

$$\Delta_0^+ + \Delta_0^- = D - 2\omega_e + \Delta_0^- - \Delta_{-1}^-, \quad (A.8)$$

which can be attained at a different magnetic field $B = B_m^{(\beta)}$. In this case, states $|-1/2, \downarrow, 0, +1/2, \uparrow, 0\rangle$ and $|-1/2, \uparrow, -1, -1/2, \downarrow, 0\rangle$ are degenerate, and the coupling matrix element is the same as in Eq. (A.7). Note, however, this energy-matching condition corresponds to a different magnetic field $B = B_m^{(\beta)}$ as Eq. (A.6) is not equivalent to (A.8).

## APPENDIX B: DETERMINING $\Gamma_{eff}$

Here we derive an expression for the effective nuclear spin polarization transport rate in the simultaneous presence of spin pumping and electronic spin-lattice relaxation. Throughout this derivation, we set the optical pumping rate to the optimum, $\Gamma_{op} = \mathcal{J}_{eff}$ and assume that $\mathcal{J}_{eff}, \Gamma_{op} > \Gamma_e$. For the particular case of the ordered chain shown in stage $(i)$ of Fig. 3b, three consecutive spins $S'$ must have the specific projections $m_{S'_{j-2}} = -1/2$, $m_{S'_{j-1}} = +1/2$, $m_{S'_j} = -1/2$ in order to enable the desired nuclear $(j-2, j-1)$ flip-flop. Under this condition, the state of two adjacent nuclear spins of opposite polarization evolves as

$$\langle \uparrow\downarrow|\rho_n|\uparrow\downarrow\rangle \sim \exp(-\Gamma_{op}t) \quad (B.1)$$
$$\langle \downarrow\uparrow|\rho_n|\downarrow\uparrow\rangle \sim 1 - \exp(-\Gamma_{op}t), \quad (B.2)$$

which essentially corresponds to the dynamics shown in Fig. 2b. It is important to emphasize that any other configuration for the three contiguous electronic spins precludes the magnetization exchange between the pair of nuclear spins.

Now we consider a more general case, and assume that at time $t = 0$, all possible configurations for these three $S'$ electronic spins are equally likely. For the sake of simplicity, we assume that the longitudinal relaxation of $S'$ spins occurs deterministically, once at the end of every interval of duration $\Gamma_e^{-1}$. We start with the occupation of the two possible nuclear configurations being $\langle\uparrow\downarrow|\rho_n|\uparrow\downarrow\rangle = 1$ and $\langle\downarrow\uparrow|\rho_n|\downarrow\uparrow\rangle = 0$. So, at time $t = \Gamma_e^{-1}$ only one out of $2^3 = 8$ configurations would have the nuclear flip-flop,

$$\langle\uparrow\downarrow|\rho_n|\uparrow\downarrow\rangle = 1 - \frac{1}{8}\left[1 - \exp\left(-\frac{\Gamma_{op}}{\Gamma_e}\right)\right] \approx 1 - \frac{1}{8} = \frac{7}{8}, \quad (B.3)$$

where we make use of the condition $\Gamma_{op}/\Gamma_e > 1$. Extending the same argument, at time $t = 2\Gamma_e^{-1}$,

$$\langle\uparrow\downarrow|\rho_n|\uparrow\downarrow\rangle = \frac{7}{8} - \frac{7}{8}\times\frac{1}{8}\left[1 - \exp\left(-\frac{\Gamma_{op}}{\Gamma_e}\right)\right] \approx \left(\frac{7}{8}\right)^2. \quad (B.4)$$

|  | $\left\|-\frac{1}{2}\uparrow 0\uparrow+\frac{1}{2}0\right\rangle$ | $\left\|-\frac{1}{2}\uparrow 0+\frac{1}{2}\downarrow 0\right\rangle$ | $\left\|-\frac{1}{2}\downarrow 0+\frac{1}{2}\uparrow 0\right\rangle$ | $\left\|-\frac{1}{2}\downarrow 0+\frac{1}{2}\downarrow 0\right\rangle$ | $\left\|-\frac{1}{2}\uparrow -1-\frac{1}{2}\uparrow 0\right\rangle$ | $\left\|-\frac{1}{2}\uparrow -1-\frac{1}{2}\downarrow 0\right\rangle$ | $\left\|-\frac{1}{2}\downarrow -1-\frac{1}{2}\uparrow 0\right\rangle$ | $\left\|-\frac{1}{2}\downarrow -1-\frac{1}{2}\downarrow 0\right\rangle$ |
|---|---|---|---|---|---|---|---|---|
| $\left\langle-\frac{1}{2}\uparrow 0+\frac{1}{2}\uparrow 0\right\|$ | $\Delta_0^+ - \Delta_0^-$ | 0 | 0 | 0 | $\mathcal{J}_d c_2 c_1$ | $-\mathcal{J}_d s_2 c_1$ | $\mathcal{J}_d c_2 s_1$ | $-\mathcal{J}_d s_2 s_1$ |
| $\left\langle-\frac{1}{2}\uparrow 0+\frac{1}{2}\downarrow 0\right\|$ | 0 | $-\Delta_0^+ - \Delta_0^-$ | 0 | 0 | $\mathcal{J}_d s_2 c_1$ | $\mathcal{J}_d c_2 c_1$ | $\mathcal{J}_d s_2 s_1$ | $\mathcal{J}_d c_2 s_1$ |
| $\left\langle-\frac{1}{2}\downarrow 0+\frac{1}{2}\uparrow 0\right\|$ | 0 | 0 | $\Delta_0^+ + \Delta_0^-$ | 0 | $-\mathcal{J}_d c_2 s_1$ | $\mathcal{J}_d s_2 s_1$ | $\mathcal{J}_d c_2 c_1$ | $-\mathcal{J}_d s_2 c_1$ |
| $\left\langle-\frac{1}{2}\downarrow 0+\frac{1}{2}\downarrow 0\right\|$ | 0 | 0 | 0 | $-\Delta_0^+ + \Delta_0^-$ | $-\mathcal{J}_d s_2 s_1$ | $-\mathcal{J}_d c_2 s_1$ | $\mathcal{J}_d s_2 c_1$ | $\mathcal{J}_d c_2 c_1$ |
| $\left\langle-\frac{1}{2}\uparrow -1-\frac{1}{2}\uparrow 0\right\|$ | $\mathcal{J}_d c_2 c_1$ | $\mathcal{J}_d s_2 c_1$ | $-\mathcal{J}_d c_2 s_1$ | $-\mathcal{J}_d s_2 s_1$ | $2\delta_\omega - \Delta_0^- - \Delta_{-1}^-$ | 0 | 0 | 0 |
| $\left\langle-\frac{1}{2}\uparrow -1-\frac{1}{2}\downarrow 0\right\|$ | $-\mathcal{J}_d s_2 c_1$ | $\mathcal{J}_d c_2 c_1$ | $\mathcal{J}_d s_2 s_1$ | $-\mathcal{J}_d c_2 s_1$ | 0 | $2\delta_\omega + \Delta_0^- - \Delta_{-1}^-$ | 0 | 0 |
| $\left\langle-\frac{1}{2}\downarrow -1-\frac{1}{2}\uparrow 0\right\|$ | $\mathcal{J}_d c_2 s_1$ | $\mathcal{J}_d s_2 s_1$ | $\mathcal{J}_d c_2 c_1$ | $\mathcal{J}_d s_2 c_1$ | 0 | 0 | $2\delta_\omega - \Delta_0^- + \Delta_{-1}^-$ | 0 |
| $\left\langle-\frac{1}{2}\downarrow -1-\frac{1}{2}\downarrow 0\right\|$ | $-\mathcal{J}_d s_2 s_1$ | $\mathcal{J}_d c_2 s_1$ | $-\mathcal{J}_d s_2 c_1$ | $\mathcal{J}_d c_2 c_1$ | 0 | 0 | 0 | $2\delta_\omega + \Delta_0^- + \Delta_{-1}^-$ |

**Table A.1**: Matrix representation for the Hamiltonian $H^{(1,2)}$ in a chain of dressed nuclei. A global ½ factor has been omitted. Notation for states corresponds to $\left|m_{S_1'}, m_{I_1}, m_{S_1}, m_{S_2'}, m_{I_2}, m_{S_2}\right\rangle$, and elaborating on Eqns. (A.3-A.5) we define $c_1 = \cos[(\theta_0^- - \theta_{-1}^-)/2]$, $s_1 = \sin[(\theta_0^- - \theta_{-1}^-)/2]$, $c_2 = \cos[(\theta_0^- - \theta_0^+)/2]$, and $s_2 = \sin[(\theta_0^- - \theta_0^+)/2]$. We introduced also the detuning factor $\delta_\omega = D - 2\omega_e$. Here we are limited to the subspaces spanned by states $\left|m_{S_1} = 0, m_{S_2'} = +1/2\right\rangle$ and $\left|m_{S_1} = -1, m_{S_2'} = -1/2\right\rangle$ for the pair $(S_1, S_2')$, denoted by green- and grey-shaded blocks, respectively. In addition, we restrict to $m_{S_1'} = -1/2$, $m_{S_2} = 0$ for the dummy (fixed) spin variables corresponding to $S_1'$ and $S_2$. Yellow-shaded matrix elements correspond to the effective coupling elements defining $\mathcal{J}_{\text{eff}}$.

Generalizing to $t = n\Gamma_e^{-1}$, we write

$$\langle\uparrow\downarrow|\rho_n|\uparrow\downarrow\rangle \approx \left(\frac{7}{8}\right)^n = \exp\left[\ln\left(\frac{7}{8}\right)\Gamma_e t\right], \quad (B.5)$$

which we use to propose an effective nuclear spin transport rate constant

$$\Gamma_{\text{eff}} \approx \left|\ln\left(\frac{7}{8}\right)\right|\Gamma_e. \quad (B.6)$$

While this value is appropriate for large chains or rings, the logarithmic pre-factor slightly changes in small systems since the fluctuations in the $S'$ electronic spins become comparatively more 'restricted' and we need to consider the configurations of only two, not three, electron spins $S'$. In this special case, the above arguments can be adapted to yield

$$\Gamma_{\text{eff}} = \left|\ln\left(\frac{3}{4}\right)\right|\Gamma_e. \quad (B.7)$$

**ACKNOWLEDGEMENTS.**

We thank Sriram Ganeshan for helpful discussions. S.B., P.R.Z. and R.H.A. acknowledge financial support from CONICET (PIP-111122013010074 6CO), SeCyT-UNC (33620180100154CB) and ANPCYT (PICT-2014-1295). Work by C.A.M. was supported by the U.S. Department of Energy (DOE), Office of Science, Basic Energy Sciences (BES) under Award BES-DE-SC0020638.

---

[1] H. Hodaei, A.U. Hassan, S. Wittek, H. Garcia-Gracia, R. El-Ganainy, D. N. Christodoulides, M. Khajavikhan, "Enhanced sensitivity at higher-order exceptional points", *Nature* **548**, 187 (2017).

[2] L. Feng, Z.J. Wong, R.-M. Ma, Y. Wang, X. Zhang, "Single-mode laser by parity-time symmetry breaking", *Science* **346**, 972 (2014).

[3] H. Hodaei, M.-A. Miri, M. Heinrich, D. N. Christodoulides, M. Khajavikhan, "Parity-time-symmetric microring lasers", *Science* **346**, 975 (2014).

[4] A. Regensburger, C. Bersch, M.-A. Miri, G. Onishchukov, D.N. Christodoulides, U. Peschel, "Parity-time synthetic photonic lattices", *Nature* **488**, 167 (2012).

[5] A. Guo, G.J. Salamo, D. Duchesne, R. Morandotti, M. Volatier-Ravat, V. Aimez, G.A. Siviloglou, D.N. Christodoulides, "Observation of PT-symmetry breaking in complex optical potentials", *Phys. Rev. Lett.* **103**, 093902 (2009).

[6] S. Weidemann, M. Kremer, T. Helbig, T. Hofmann, A. Stegmaier, M. Greiter, R. Thomale, A. Szameit, "Topological funneling of light", *Science* **368**, 311 (2020).

[7] B. Peng, S. K. Özdemir, F. Lei, F. Monifi, M. Gianfreda, G. L. Long, S. Fan, F. Nori, C. M. Bender, L. Yang, "Nonreciprocal light transmission in parity-time-symmetric whispering-gallery microcavities", *Nat. Phys.* **10**, 394 (2014).

[8] L. Chang, X. Jiang, S. Hua, C. Yang, J. Wen, L. Jiang, G. Li, G. Wang, M. Xiao, "Parity-time symmetry and variable optical isolation in active-passive-coupled microresonators", *Nat. Phot.* **8**, 524 (2014).

[9] A. B. Khanikaev, A. Alù, "Optical isolators: Nonlinear dynamic reciprocity", *Nat. Phot.* **9**, 359 (2015).

[10] L.D. Tzuang, K. Fang, P. Nussenzveig, S. Fan, M. Lipson, "Non-reciprocal phase shift induced by an effective magnetic flux for light", *Nat. Photon.* **8**, 701 (2014).

[11] E. Li, B. J. Eggleton, K. Fang, S. Fan, "Photonic Aharonov-Bohm effect in photon-phonon interactions", *Nat. Commun.* **5**, 3225 (2014).

[12] K. Fang, Z. Yu, S. Fan, "Realizing effective magnetic field for photons by controlling the phase of dynamic modulation", *Nat. Phot.* **6**, 782 (2012).

[13] R. Fleury, D. L. Sounas, C. F. Sieck, M. R. Haberman, A. Alù, "Sound isolation and giant linear nonreciprocity in a compact acoustic circulator", *Science* **343**, 516 (2014).

[14] Y. Wu, W. Liu, J. Geng, X. Song, X. Ye, C-K. Duan, X. Rong, J. Du, "Observation of parity-time symmetry breaking in a single-spin system", *Science* **364**, 878 (2019).

[15] B. Cowan, *Nuclear Magnetic Resonance and Relaxation*, Cambridge University Press, Cambridge, 2005.

[16] B. Corzilius, "High-field dynamic nuclear polarization", *Ann. Rev. Phys. Chem.* **71**, 143 (2020).

[17] D. Wisser, G. Karthikeyan, A. Lund, G. Casano, H. Karoui, M. Yulikov, G. Menzildjian, A.C. Pinon, A. Purea, F. Engelke, S.R. Chaudhari, D. Kubicki, A.J. Rossini, I.B. Moroz, D. Gajan, C. Copéret, G. Jeschke, M. Lelli, L. Emsley, A. Lesage, O. Ouari, "BDPA-nitroxide biradicals tailored for efficient dynamic nuclear polarization enhanced solid-state NMR at magnetic fields up to 21.1 T", *J. Am. Chem. Soc.* **140**, 13340 (2018).

[18] G. Stevanato, D.J. Kubicki, G. Menzildjian, A-S. Chauvin, K. Keller, M. Yulikov, G. Jeschke, M. Mazzanti, L. Emsley, "A factor two improvement in high-field dynamic nuclear polarization from Gd(III) complexes by design", *J. Am. Chem. Soc.* **141**, 8746 (2019).

[19] A. Equbal, K. Tagami, S. Han, "Balancing dipolar and exchange coupling in biradicals to maximize cross effect dynamic nuclear polarization", *Phys. Chem. Chem. Phys.* **22**, 13569 (2020).

[20] F. Mentink-Vigier, "Optimizing nitroxide biradicals for cross-effect MAS-DNP: the role of g-tensors' distance", *Phys. Chem. Chem. Phys.* **22**, 3643 (2020).

[21] M.S. Fataftah, D.E. Freedman, "Progress towards creating optically addressable molecular qubits", *Chem. Commun.* **54**, 13773 (2018).

[22] S.L. Bayliss, D.W. Laorenza, P.J. Mintun, B. Diler, D.E. Freedman, D.D. Awschalom, "Optically addressable molecular spins for quantum information processing", *Science* **370**, 1309 (2020).

[23] F. Luis, A. Repollés, M. J. Martínez-Pérez, D. Aguilà, O. Roubeau, D. Zueco, P.J. Alonso, M. Evangelisti, A. Camón, J. Sesé, L.A. Barrios, G. Aromí, "Molecular prototypes for spin-based cnot and swap quantum gates", *Phys. Rev. Lett.* **107**, 117203 (2011).

[24] M. Atzori, R. Sessoli, "The second quantum revolution: Role and challenges of molecular chemistry", *J. Am. Chem. Soc.* **141**, 11339 (2019).

[25] D. Aguilà, L.A. Barrios, V. Velasco, O. Roubeau, A. Repollés, P.J. Alonso, J. Sesé, S.J. Teat, F. Luis, G. Aromí, "Heterodimetallic [LnLn′] Lanthanide Complexes: Toward a chemical design of two-qubit molecular spin quantum gates", *J. Am. Chem. Soc.* **136**, 14215 (2014).

[26] F. Lombardi, A. Lodi, J. Ma, J. Liu, M. Slota, A. Narita, W.K. Myers, K. Müllen, X. Feng, L. Bogani, "Quantum units from the topological engineering of molecular graphenoids", *Science* **366**, 1107 (2019).

[27] M. Atzori, A. Chiesa, E. Morra, M. Chiesa, L. Sorace, S. Carretta, R. Sessoli, "A Two-qubit molecular architecture for electron-mediated nuclear quantum simulation", *Chem. Sci.* **9**, 6183 (2018).

[28] D. Pagliero, K.R. Koteswara Rao, P.R. Zangara, S. Dhomkar, H.H. Wong, A. Abril, N. Aslam, A. Parker, J. King, C.E. Avalos, A. Ajoy, J. Wrachtrup, A. Pines, C.A. Meriles, "Multispin-assisted optical pumping of bulk $^{13}$C nuclear spin polarization in diamond", *Phys. Rev. B* **97**, 024422 (2018).

[29] D. Pagliero, P. Zangara, J. Henshaw, A. Ajoy, R.H. Acosta, J.A. Reimer, A. Pines, C.A. Meriles, "Optically pumped spin polarization as a probe of many-body thermalization", *Science Adv.* **6**, eaaz6986 (2020).

[30] D. Pagliero, P.R. Zangara, J. Henshaw, A. Ajoy, R.H. Acosta, N. Manson, J. Reimer, A. Pines, C.A. Meriles, "Magnetic-field-induced delocalization in hybrid electron-nuclear spin ensembles", *Phys. Rev. B* **103**, 064310 (2021).

[31] J.R. Johansson, P.D. Nation, F. Nori, "QuTiP 2: A Python framework for the dynamics of open quantum systems", *Comput. Phys. Commun.* **184**, 1234 (2013).

[32] H. Carmichael, *An open systems approach to quantum optics*, Springer-Verlag, Berlin, 1993.

[33] C. Gardiner, P. Zoller, *Quantum noise: a handbook of Markovian and non-Markovian quantum stochastic methods with applications to quantum optics*, Springer Science & Business Media, Berlin, 2004.

[34] X. Zotos, F. Naef, P. Prelovsek, "Transport and conservation laws", *Phys. Rev. B* **55**, 11029 (1997).

[35] L. Schuab, E. Pereira, G.T. Landi, "Energy rectification in quantum graded spin chains: Analysis of the XXZ model", *Phys. Rev. E* **94**, 042122 (2016).

[36] A. De Luca, M. Collura, J. De Nardis, "Nonequilibrium spin transport in integrable spin chains: Persistent currents and emergence of magnetic domains", *Phys. Rev. B* **96**, 020403(R) (2017).

[37] J. Henshaw, D. Pagliero, P.R. Zangara, B. Franzoni, A. Ajoy, R. Acosta, J.A. Reimer, A. Pines, C.A. Meriles, "$^{13}$C dynamic nuclear polarization in diamond via a microwave-free 'integrated' cross effect", *Proc. Natl. Acad. Sci. USA* **116**, 18334 (2019).

[38] P.R. Zangara, J. Henshaw, D. Pagliero, A. Ajoy, J.A. Reimer, A. Pines, C.A. Meriles, "Two-electron-spin ratchets as a platform for microwave-free dynamic nuclear polarization of arbitrary material targets", *Nano Lett.* **19**, 2389 (2019).

[39] Z. Denis, A. Biella, I. Favero, C. Ciuti, "Permanent directional heat currents in lattices of optomechanical resonators", *Phys. Rev. Lett.* **124**, 083601 (2020).

[40] T.E. Lee, "Anomalous edge state in a non-Hermitian lattice", *Phys. Rev. Lett.* **116**, 133903 (2016).

[41] V.M. Martinez Alvarez, J.E. Barrios Vargas, M. Berdakin, L.E.F. Foa Torres, "Topological states of non-Hermitian systems", *Eur. Phys. J. - Spec. Top.* **227**, 1295 (2018).

[42] S. Yao, Z. Wang, "Edge states and topological invariants of non-Hermitian systems", *Phys. Rev. Lett.* **121**, 086803 (2018).

[43] C.H. Lee, R. Thomale, "Anatomy of skin modes and topology in non-Hermitian systems", *Phys. Rev. B* **99**, 201103 (2019).

[44] A. Ghatak, M. Brandenbourger, J. van Wezel, C. Coulais, "Observation of non-Hermitian topology and its bulk-edge correspondence", A. Ghatak, M. Brandenbourger, J. van Wezel, C. Coulais, arXiv 1907.11619 (2019).

[45] T. Helbig, T. Hofmann, S. Imhof, M. Abdelghany, T. Kiessling, L.W. Molenkamp, C.H. Lee, A. Szameit, M. Greiter, R. Thomale, "Observation of bulk boundary correspondence breakdown in topolectrical circuits", *Nat. Phys.* **16**, 747 (2020).

[46] L. Xiao, T. Deng, K. Wang, G. Zhu, Z. Wang, W. Yi, P. Xue, "Non-Hermitian bulk–boundary correspondence in quantum dynamics", *Nat. Phys.* **16**, 761 (2020).

[47] I. Gilary, A.A. Mailybaev, N. Moiseyev, "Time-asymmetric quantum-state-exchange mechanism", *Phys. Rev. A* **88**, 010102(R) (2013).